\documentclass[aps,prd,preprint,tightenlines,superscriptaddress]{revtex4}
\usepackage{epsfig}
\usepackage{graphicx}
\usepackage{psfrag}
\usepackage{amsmath,amssymb}
\usepackage{colordvi}
\usepackage{color}
\usepackage{amsfonts}
\usepackage{enumerate}
\usepackage{slashed,bm}

\begin{document}

\title{Physics of Gluon Helicity Contribution to Proton Spin}
\author{Xiangdong Ji}
\affiliation{INPAC, Department of Physics and Astronomy, Shanghai Jiao Tong University, Shanghai, 200240, P. R. China}
\affiliation{Center for High-Energy Physics, Peking University, Beijing, 100080, P. R. China}
\affiliation{Maryland Center for Fundamental Physics, University of Maryland, College Park, Maryland 20742, USA}
\author{Jian-Hui Zhang}
\affiliation{INPAC, Department of Physics and Astronomy, Shanghai Jiao Tong University, Shanghai, 200240, P. R. China}
\author{Yong Zhao}
\affiliation{Maryland Center for Fundamental Physics, University of Maryland, College Park, Maryland 20742, USA}

\date{\today}
\vspace{0.5in}
\begin{abstract}

The total gluon helicity in a polarized proton, measurable in high-energy scattering,
is shown to be the large momentum limit of a gauge-invariant but
non-local, frame-dependent gluon spin $\vec{E}\times \vec{A}_\perp$ in QCD. This
opens a door for a non-perturbative calculation of this quantity in lattice QCD, and also 
justifies using free-field expressions in the light-cone gauge as physical observables. 
 
\end{abstract}

\maketitle

The total gluon helicity $\Delta G=\int dx\Delta g(x)$ in a longitudinally polarized
proton is an important physical quantity that has motivated much experimental effort to
measure in high-energy scattering~\cite{compass, hermes, phenix, star}. It helps to understand,
among others, how the helicity of a fast-traveling proton is composed of partons'
helicity and orbital angular momentum~\cite{Jaffe:1989jz, brodsky:1995, jaffe:1996, xji:1996, Chen:2006ng}.
The factorization theorems in quantum chromodynamics (QCD) indicate that $\Delta G$ is a matrix element of
a complicated light-cone correlation operator of the gluon fields~\cite{manohar}, and
has a simple physical interpretation only in the light-cone
gauge $A^+=0$ natural for parton physics~\cite{Jaffe:1989jz}. This state of understanding
makes computing $\Delta G$ in lattice QCD infeasible, and raises the fundamental question
about the gauge invariance of the 
gauge particle spin in a bound state~\cite{Jackson}.

In this paper, we report a breakthrough in understanding the physics
of $\Delta G$, and correspondingly leading to a practical way to its calculation
in lattice QCD. We find that $\Delta G$ can be obtained by boosting a matrix element
of the gluon spin operator---$\vec{E}\times \vec{A}_\perp$---to
the infinite momentum frame (IMF), where $\vec{A}_\perp$ is the transverse part of the gauge potential.
This operator was first proposed in Ref.~\cite{Chen:2008ag} as the gauge-invariant gluon spin, 
but has been criticized as physically uninteresting because of its frame dependence~\cite{Ji:2010zza}.
The physics behind the IMF limit we propose here goes back to the well-known Weiss\"{a}cker-Williams equivalent photon picture
for high-energy scattering~\cite{Jackson}. But of course, there are subtleties in
taking the IMF limit. In particular, the matrix element has a singular dependence
on the bound state momentum in perturbation theory as it approaches infinity. Moreover,
the anomalous dimension of $\vec{E}\times \vec{A}_\perp$ does not coincide with that of the non-local
light-cone correlation in the factorization theorems~\cite{chen:ag2011}. Therefore, we will provide a
well-defined procedure, or a matching condition, for the limiting procedure. In particular, we will show in a one-loop example
how to obtain $\Delta G$ from the IMF limit of a frame-dependent, time-independent
matrix element. This example helps to demonstrate that a non-perturbative $\Delta G$
can be recovered from a frame-dependent lattice matrix element of an Euclidean space operator
$\vec{E}\times \vec{A}_\perp$.\\

The difficulty in understanding and calculating $\Delta G$ is easy to appreciate. Through QCD
factorization, it has been shown that $\Delta G$ is a matrix element of a
non-local operator involving light-cone correlation~\cite{manohar},
\begin{equation}
  \Delta G = \int dx \frac{i}{2xP^+} \int
  \frac{d\xi^-}{2\pi} e^{-ixP^+\xi^-} \langle PS| F^{+\alpha}_a(\xi^-) {\cal L}^{ab}(\xi^-,0)\tilde F_{\alpha,b}^{~+} (0)|PS\rangle \ ,
\label{gpdf}
\end{equation}
where  $|PS\rangle$ is a proton plane-wave state with momentum $P^\mu$ and polarization $S^\mu$, $\tilde F^{\alpha\beta} = (1/2)\  \epsilon^{\alpha\beta\mu\nu}F_{\mu\nu}$,
and ${\cal L}(\xi^- ,0) = P\exp[-ig\int^{\xi^-}_0  \mathcal{A}^+(\eta^-,0_\perp)\ d\eta^-]$ with
$\mathcal{A}^+ \equiv T^cA^+_c$ is a light-cone gauge link defined in the adjoint representation. The light-front coordinates are defined as $\xi^\pm = (\xi^0\pm \xi^3)/\sqrt{2}$. It is usually difficult to see
the above operator as the gluon spin or helicity. However, in the light-cone
gauge, $A^+=0$, the whole operator collapses into $\vec{E}\times \vec{A}$---the textbook definition
of the gauge particle spin~\cite{Jackson}---which is known to be gauge dependent, and $\Delta G$ can be regarded as the number of gluon partons with helicity
in the direction of the proton helicity minus that with opposite helicity. Because of the
explicit presence of the real time in $\xi^-$, one cannot evaluate the
above expression in lattice QCD. An early attempt to get the gluon helicity on lattice was to calculate the
matrix element of $F_{\mu\nu}\tilde F^{\mu\nu}$~\cite{mandula}, but there is no demonstrated
connection between this and $\Delta G$.

To find the physics of $\Delta G$ without committing to the light-cone gauge, let us examine the operator structure a bit further. For simplicity, we first consider the U(1) gauge theory (quantum electrodynamics, or QED) so that the gauge link is absent. Carrying out the integration
over the longitudinal momentum, the gauge-invariant photon ``spin" operator becomes:
\begin{align}
\hat S_\gamma^{\rm inv}(0)&={i}\int\!\! \frac{dx}{x}\!\! \int\!\! \frac{d^2 k_\perp}{(2\pi)^3} \int\!\! {d\xi^- d^2\xi_\perp} e^{-i(xP^+\xi^- - \vec k_\perp\cdot\vec\xi_\perp)}
\left[ixP^+A^i(\xi^-,\vec\xi_\perp)-ik^i_\perp A^+(\xi^-,\vec\xi_\perp)\right]\tilde F_{i}^{~+} (0) \nonumber\\
&= - \int\frac{dk^+ d^2k_\perp}{(2\pi)^3} \Big[\tilde{A}^i(k^+,\vec k_\perp)
-\frac{k^i_\perp}{k^+}\tilde A^+(k^+,\vec k_\perp)\Big]\tilde F_{i}^{~+} (0)
\nonumber \\
&= \Big[\vec{E}(0)\times \Big(\vec{A}(0) -  \frac{1}{\nabla^+}\vec{\nabla}A^+(\xi^-)\Big) \Big]^3 \ ,
\label{photonspin}
\end{align}
where $\nabla^+ = \partial/\partial\xi^-$, $E^i = F^{i+}$, $k^+=xP^+$, and the third component of a vector is interpreted in
the usual sense of the cross product. The $\xi^-$ coordinate in $A^+$ is taken to 0 after operating with
the inverse derivative.

The above operator is just the IMF limit of $\vec{E}\times \vec{A}_\perp$, where $E^i=F^{i0}$,
and  $A^i_\perp = (\delta^{ij} - \nabla^i\nabla^j/\nabla^2) A^j$
is the transverse part of the gauge potential and is invariant under gauge transformation.
The rule of taking the IMF limit of an operator is as follows:
For any vector $V^\mu$, define $V^\pm = (V^0\pm V^3)/\sqrt{2}$. If the boost is along the 3-direction,
then the components of the vector go like $V^+\rightarrow V^+\Lambda$, $V^-\rightarrow V^-/\Lambda$,
and $V_\perp \rightarrow V_\perp$. Thus $\nabla^2\rightarrow (\nabla^+)^2\Lambda^2$ and $\vec{\nabla}\cdot \vec{A}\rightarrow
\nabla^+A^+\Lambda^2$ for the leading components. Using these rules, one finds that
$\vec{E}\times \vec{A}_\perp \rightarrow  \vec{E}\times (\vec{A}- (1/\nabla^+)\vec{\nabla}A^+)$,
which is exactly the above operator in Eq.~(\ref{photonspin}).

It is known that in electromagnetic theory the vector potential can be uniquely
separated into the longitudinal and transverse parts, $\vec{A}=\vec{A}_{\parallel} + \vec{A}_\perp$,
and the transverse part is gauge invariant~\cite{belinfante}: Given $\vec{A}$, $\vec{A}_\perp$
can be uniquely constructed as a functional of $\vec{A}$ with an appropriate boundary condition.
Thus $\vec{E}\times \vec{A}_\perp$ is a gauge-invariant operator, and can be regarded as the
gauge-invariant part of the gauge particle spin~\cite{Chen:2008ag}. $\vec{E}\times \vec{A}_\perp$ is
a non-local operator in that it depends on the gauge potential over all space.

It is important to realize that separating $\vec{A}$ and $\vec{E}$ into longitudinal and transverse
parts is in general not a physically meaningful thing to do. In the first place, the physics of $\vec{E}$
is to apply force to electric charge and there is no charge that responds separately to $\vec{E}_{\parallel }$
and $\vec{E}_\perp$. Second, in a different frame, one sees different transverse and longitudinal
separations, and therefore the notion has no Lorentz covariance~\cite{Ji:2010zza}. As we shall see, the frame-dependence of both parts
is dynamical, and cannot be calculated without solving the theory. However, there are two exceptions
where the separation is meaningful. The first case concerns the radiation field~\cite{Ji:2012gc}. For free radiation,
by separating out the unphysical degrees of freedom, one simplifies the quantization procedure
significantly. In the laser beam, this separation allows one to talk about the gauge-invariant photon spin
and orbital angular momentum~\cite{Allen1992}. The second case is the IMF, which is our interest here.
In the IMF, $E_{\parallel }\ll E_{\perp}$, and the electromagnetic field can be regarded as free radiation. This has been recognized long ago by Weiss\"acker and Williams, in the name of equivalent photon approximation~\cite{Jackson}.
Only then, $\vec{E}\times \vec{A}_\perp$ can be understood as a physical quantity, where $\vec{E}$ can also
be replaced by $\vec{E}_{\perp }$.

Therefore, the photon helicity measurable in high-energy scattering can be calculated as the IMF
limit of a matrix element of the static operator $\vec{E}\times \vec{A}_\perp$. To calculate the matrix element of the time-independent, albeit non-local,
operator is a standard practice in lattice QCD. It will be dynamically dependent on the
momentum of the external particle~\cite{Ji:2010zza}. In fact, the dependence is
singular in the leading order perturbation theory: In the IMF limit,
the matrix element diverges. To obtain the physically interesting
finite light-cone matrix element, one has to find a matching condition, which we
will come to after establishing a similar connection in QCD.

The case for QCD is a bit more complicated. Separating $\vec{A}$ into longitudinal and transverse
parts requires generalizing the observations in QED to similar conditions in QCD, which has been considered
long ago~\cite{treat:1972} (see also Ref.~\cite{Chen:2008ag}). Clearly, we would like
to have $\vec{A}_\perp$ transform covariantly under gauge transformation:
\begin{equation}
           \vec{{A}}_\perp \rightarrow  U(x)\vec{{A}}_\perp U^\dagger(x)  \ ,
\end{equation}
where $\vec{{A}}_\perp \equiv T_a\vec{A}^a_\perp$, so it is easy to construct gauge-invariant quantities with $\vec{A}_\perp$.
Second, we require $\vec{A}_{\parallel }$ to produce null magnetic field, as it does in QED.
This condition is~\cite{treat:1972}:
\begin{equation}\label{nullfieldstrength}
     \partial^i A^{j,a}_{\parallel} - \partial^j A^{i,a}_{\parallel} -gf^{abc}A_{\parallel}^{i,b}A_{\parallel}^{j,c} \ =\ 0 \ ,
\end{equation}
which is a nonlinear equation to solve for $A^i_{\parallel}$ as a functional of $A^i$. Moreover, the transverse part of the gauge potential satisfies a generalized Coulomb condition~\cite{treat:1972}:
\begin{equation}
\partial^i {A}^i_\perp \ = \ ig[{A}^i, {A}^i_\perp] \ .
\label{coulomb}
\end{equation}

We can then go through a similar derivation as in the QED case and find:
\begin{equation}
\hat S_g^{\rm inv}(0) = \Big[\vec{E}^a(0) \times \Big(\vec{A}^a(0) -\frac{1}{\nabla^+} (\vec{\nabla}A^{+,b}){\cal L}^{ba}(\xi^-,0)\Big) \Big]^3 \ ,
\label{gluonspin}
\end{equation}
where the inverse derivative acts on everything after it, and takes the $\xi^-$ coordinate
in the gauge link to 0.

It is a bit involved to show that the expression in the parenthesis above is indeed $\vec{A}_\perp$ in the IMF.
Since $\vec{A}_\perp = \vec{A} - \vec{A}_{\parallel }$, we just need to solve for $\vec{A}_{\parallel }$. 
After solving Eq.~(\ref{nullfieldstrength}) and Eq.~(\ref{coulomb}) order by order in $g$, we find that $A^+_\perp$ vanishes, and thus $A^+_\parallel=A^+$ in the IMF. Substituting this into Eq.~(\ref{nullfieldstrength}), we obtain a first-order
inhomogeneous linear equation for $A^i_{\parallel}$:
\begin{equation}
          \partial^+ A^{i,a}_{\parallel} -gf^{abc}A^{+,b}A_{\parallel}^{i,c} \ =\  \partial^i A^{+,a}\ .
\label{nullm}
\end{equation}
Its solution is easy to construct as a geometric series expansion:
\begin{eqnarray}
               A^{i,a}_{\parallel } &=& \frac{1}{\nabla^+}\left[1 + \left(-ig \mathcal{A}^+\frac{1}{\nabla^+}\right) + ...
+\left(-ig \mathcal{A}^+\frac{1}{\nabla^+}\right)^n + ...\right]^{ab}(\partial^i A^{+,b}) \ .
\end{eqnarray}
By commuting $\partial^i A^+$ systematically to the front of the expression, one finds:
\begin{equation}\label{Aparasolution}
 A^{i,a}_{\parallel }(\xi^-) = \frac{1}{\nabla^+}\Big((\partial^i A^{+,b}){\cal L}^{ba}(\xi'^-,\xi^-)\Big) \ ,
\end{equation}
where the coordinate $\xi'^-$ in the gauge link is taken to $\xi^-$ after operating with the inverse derivative.

Alternatively, one can multiply a gauge link $\mathcal L$ on both sides of Eq.~(\ref{nullm}), and find after some manipulations:
\begin{equation}
\partial^+(A^{i,a}_{\parallel}\mathcal L^{ad})=(\partial^i A^{+,a})\mathcal L^{ad}\ .
\end{equation}
It is then straightforward to see that the solution for $A^{i, a}_\parallel$ is formally given by Eq.~(\ref{Aparasolution}).
Clearly $\vec{A}^{ a}_{\parallel}(0)$ is just the part subtracted from $\vec{A}^{a}(0)$ in the gluon helicity operator in Eq.~(\ref{gluonspin}).
Therefore, we established the same conclusion in QCD that the gluon helicity operator
is the IMF limit of the gauge-invariant, non-local gluon spin operator $(\vec{E}\times \vec{A}_\perp)^3$.\\

\begin{figure}[hbt]
\begin{center}
\includegraphics[width=30mm]{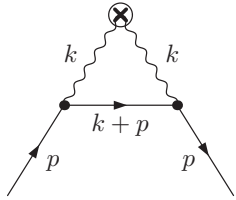}
\caption{Matrix element of $\vec{E}\times\vec{A}_\perp$ in an asymptotic electron state.}
\label{phqcorrection}
\end{center}
\end{figure}

Now we show that the matrix element of $\hat S^{\text{inv}}_\gamma = (\vec{E}\times \vec{A}_\perp)^3$ depends on the choice
of frames dynamically. By ``dynamically" we mean that the frame dependence cannot be obtained
from Lorentz transformation, and is a function of dynamic details. Let us consider the example of
photon spin in a free electron state $|p, s\rangle$. A simple perturbative calculation of Fig.~\ref{phqcorrection} yields,

\begin{equation}
\langle p,s|\hat{S}^{\text{inv}}_\gamma|p,s\rangle = {\alpha_\text{em}\over 4\pi}\left[{5\over3}D + {31\over9} + 2\int_0^1dx\sqrt{1-x}\ln\left(1+x{\vec{p}^{~2}\over m^2}\right)\right] \bar{u}(p,s)\Sigma^3u(p,s)\ .
\end{equation}
where $p^\mu = (p^0, 0, 0, p^3)$, $D=1/\epsilon - \gamma_E +\ln 4\pi + \ln(\mu^2/m^2)$, and $m$ is the mass of the electron used to
regularize the infrared (IR) divergence. The result has a non-trivial dependence on the electron momentum $\vec{p}^{~2}$, which makes its physical interpretation less straightforward. In particular,
the result is divergent in the IMF limit.  The ultra-violet (UV)
part of the matrix element is consistent with that found in Ref.~\cite{chen:ag2011}. Our result also shows that
the statement in Ref.~\cite{Wakamatsu:2010cb} about the frame independence is incorrect.

However, the measurable photon helicity can be obtained by the same matrix element by going to
the IMF limit first before the loop-momentum integration. In this limit, the external momentum dependence
is dropped out, and the matrix element becomes:
\begin{equation}
\langle p,s|\hat{S}^{\text{inv}}_\gamma|p,s\rangle = {\alpha_\text{em}\over 4\pi} (3D+7)\ \bar{u}(p,s)\Sigma^3u(p,s) \ .
\label{spinzimf}
\end{equation}
This is exactly the same as that computed using the factorization expression in Eq.~(\ref{gpdf}),
or using $\vec{E}\times \vec{A}$ (notice the full $\vec{A}$), in the light-cone gauge $A^+=0$~\cite{Hoodbhoy:1999dr}.
The UV property of the matrix element is the same as that derived by Altarelli and Parisi (AP)~\cite{Altarelli:1977zs}.
In fact, in the original derivation of AP evolution equation, a finite frame result was used to obtain
parton physics in the IMF limit. Our result also indicates the claim that gluons carry only about $1/5$ of the nucleon momentum in Ref.~\cite{Chen:2008ag} is incorrect, because the matrix elements of the quark and gluon momentum operators were calculated in the finite momentum frame. We have verified that if they are boosted to the IMF, we can get the standard mixing matrix in Ref.~\cite{Poli74}. Meanwhile, it should be pointed out that the result in Ref.~\cite{Wakamatsu:2012ve} is standard because they used the light-cone gauge condition.

The above calculation shows that the IMF and UV limits are not exchangeable. However, since the
IR part is the same, the difference is a perturbatively calculable quantity.
This turns out to be the key for a non-perturbative computation of the gluon helicity. Since lattice QCD
cannot handle the real time dependence, a direct calculation of Eq.~(\ref{gpdf}) is infeasible.
However, one can get the same matrix element by studying the matrix element of
$\vec{S}_g^{\rm inv} = \vec{E}\times \vec{A}_\perp$ as a function of external momentum $\vec{P}$.
The largest momentum attainable on a lattice is of order $1/a$, with $a$
being the lattice spacing. On the other hand, the matrix element also has UV dependence on $1/a$, and
can be calculated in perturbation theory. Thus we can calculate the matrix element at the
largest momentum $\sim 1/a$,
matching the results of the two different limits in a perturbative way. For instance, in the above one-loop example
with a finite $\epsilon$, one can match the two results by setting:
\begin{equation}
   \ln \vec{p}^{~2}  = (D + \ln m^2) + \frac{16}{3} - 2\ln2 \ .
\end{equation}
The result will become more accurate as the renormalization scale is asymptotically large. On the lattice,
$D+\ln m^2$ is replaced by $-\ln a^2$, and the matching condition becomes $\ln (pa)^2=$ const. We will explore the issue of the lattice calculation in more detail in a separate publication~\cite{jizhangzhao}.\\

At last, it is useful to consider the frame dependence of the
angular momentum sum rule which has recently been strongly advocated in the literature~\cite{Chen:2008ag, Wakamatsu:2010mb, Wakamatsu:2010cb}. The primary goal is to find a simple free-field form of the angular momentum decomposition
so that the individual parts have simple physical interpretation. The closest gauge-invariant form
involves the expression~\cite{Chen:2008ag}:
\begin{eqnarray}
   \vec{J} &=& \int d^3x\ \psi^\dagger \frac{1}{2}\vec{\Sigma} \psi
   + \int d^3x\ \psi^\dagger \vec{x}\times\frac{1}{i}(\vec{\nabla} - ig\vec{{A}}_\parallel)\psi\nonumber\\
\ \ \ \ &&+ \int d^3x\ \vec{E}_a\times\vec{A}^a_\perp
+ \int d^3x\ E_a^i\ \vec{x}\times\vec{\nabla} A^{i,a}_\perp \ .
\end{eqnarray}
This result is frame dependent and not physically interesting in general. However,
in the IMF one has $A_{\parallel}^{i,a}(\xi^-) = (1/\nabla^+)\big(\nabla^i A^{+, b}\mathcal{L}^{ba}(\xi'^-,\xi^-)\big)$;
the above decomposition is the same as the Jaffe-Manohar result in the light-cone gauge $A^+=0$~\cite{Jaffe:1989jz}.
Therefore, this serves to justifying that the light-cone gauge is the natural choice in the IMF, where free-field
expressions such as $\vec{E}\times \vec{A}$ attain physical significance. In particular,
the corresponding matrix elements are physically measurable.\\

To conclude, we have shown that the total gluon helicity measured in high-energy scattering
is the IMF limit of  a matrix element of a gauge-invariant operator. This limit
does not commute with the UV limit in quantum field theory and therefore
the two operators have different anomalous dimensions at first sight. However, they can be
related through a matching condition. This allows an otherwise infeasible lattice QCD
calculation of light-cone correlations. We have also explained why free-field theory expressions in the light-cone gauge
are physically meaningful, as they correspond to the IMF limit of gauge-invariant, but
non-local expressions in interacting theories.

\vspace{1em}
We thank J.~-W. Chen for useful discussions about the gauge-invariant gluon spin operator.
This work was partially supported by the U.
S. Department of Energy via grants DE-FG02-93ER-40762
and a grant (No. 11DZ2260700) from the Office of
Science and Technology in Shanghai Municipal Government,  and by the National Science Foundation of China (No. 11175114).


\begin{thebibliography}{99}

\bibitem{compass}
  The COMPASS Collaboration,
  Phys.\ Lett.\ B\ {\bf 633}, 25 (2006);
  Phys.\ Lett.\ B\ {\bf 647}, 8 (2007);
  arXiv:[hep-ex] 0802.3023 (2008);
  Phys.\ Lett.\ B\ {\bf 676}, 31 (2009);
  Phys.\ Lett.\ B\ {\bf 718}, 922 (2013);
  Phys.\ Rev.\ D {\bf 87}, 052018 (2013).


\bibitem{hermes}
  The HERMES Collaboration,
  JHEP {\bf 1008}, 130 (2010).


\bibitem{phenix}
  The PHENIX Collaboration,
  Phys.\ Rev.\ Lett.\ {\bf 93}, 202002 (2004);
  Phys.\ Rev.\ D\ {\bf 76}, 051106 (2007);
  Phys.\ Rev.\ D\ {\bf 79}, 012003 (2009);
  Phys.\ Rev.\ Lett.\ {\bf 103}, 012003 (2009);
  Phys.\ Rev.\ D\ {\bf 83}, 032001 (2011);
  Phys.\ Rev.\ D\ {\bf 84}, 012006 (2011);
  Phys.\ Rev.\ D\ {\bf 86}, 092006 (2012);
  Phys.\ Rev.\ D\ {\bf 87}, 012011 (2013).


\bibitem{star}
  The STAR Collaboration,
  Phys.\ Rev.\ Lett.\ {\bf 97}, 252001 (2006);
  Phys.\ Rev.\ Lett.\ {\bf 100}, 232003 (2008);
  Phys.\ Rev.\ D\ {\bf 80}, 111108 (2009).


\bibitem{Jaffe:1989jz}
  R.~L.~Jaffe and A.~V.~Manohar,
  Nucl.\ Phys.\ B\ {\bf 337}, 509 (1990).


\bibitem{brodsky:1995}
  S.~ Brodsky,\ M.~Burkardt,\ and I.~Schmidt,
  Nucl.\ Phys.\ B\ {\bf 441}, 197 (1995).


\bibitem{jaffe:1996}
  R.~L.~Jaffe,
  Phys.\ Lett.\ B\ {\bf 365}, 359 (1996).


\bibitem{xji:1996}
  X.~Ji, J.~Tian, and P.~Hoodbhoy,
  Phys.\ Rev.\ Lett.\ {\bf 76}, 740 (1996).


\bibitem{Chen:2006ng}
  P.~Chen and X.~Ji,
  Phys.\ Lett.\ B {\bf 660}, 193 (2008).


\bibitem{manohar}
  A.~V.~Manohar,
  Phys.\ Rev.\ Lett.\  {\bf 66}, 289 (1991).


\bibitem{Jackson}
  J.~D.~Jackson, {\it Classical Electrodynamics}, (3rd Edition, John Wiley \& Sons Inc. 1999).


\bibitem{Chen:2008ag}
  X.~-S.~Chen, X.~-F.~L\"u, W.~-M.~Sun, F.~Wang, and T.~Goldman,
  Phys.\ Rev.\ Lett.\  {\bf 100}, 232002 (2008);
  Phys.\ Rev.\ Lett.\  {\bf 103}, 062001 (2009).


\bibitem{Ji:2010zza}
  X.~Ji,
  Phys.\ Rev.\ Lett.\  {\bf 104}, 039101 (2010);
  Phys.\ Rev.\ Lett.\  {\bf 106}, 259101 (2011).


\bibitem{chen:ag2011}
  X.~-S.~Chen, W.~-M.~Sun, F.~Wang, and T.~Goldman,
  Phys.\ Lett.\ B\ {\bf 700}, 21 (2011).


\bibitem{mandula}
  J.~E.~Mandula,
  Phys.\ Rev.\ Lett.\ {\bf 65}, 1403 (1990).


\bibitem{belinfante}
  F.~J.~Belinfante,
  Phys.\ Rev.\ {\bf 128}, 2832 (1961);
  F.~Rohrlich and F.~Strocchi,
  Phys. Rev. {\bf 139}, B476 (1965);
  V.~B.~Berestetskii, E.~M.~Lifshitz, and L.~P.~Pitaevskii,
  {\it Quantum Electrodynamics}, (2nd edition, Pergamon, Oxford, 1982);
  C.~Cohen-Tannoudji, J.~Dupont-Roc, and G.~Grynberg,
  {\it Photons and Atoms}, (John Wiley \& Sons Inc. 1989).


\bibitem{Ji:2012gc}
  X.~Ji, Y.~Xu, and Y.~Zhao,
  JHEP {\bf 1208}, 082 (2012).


\bibitem{Allen1992}
  L.~Allen, M.~W.~Beijersbergen, R.~J.~C.~Spreeuw, and J.~P.~Woerdman,
  Phys.\ Rev.\ A {\bf 45}, 8185 (1992).


\bibitem{treat:1972}
  R.~P.~Treat,
  J.\ Math.\ Phys.\ {\bf 13}, 1704 (1972).


\bibitem{Wakamatsu:2010cb}
  M.~Wakamatsu,
  Phys.\ Rev.\ D {\bf 83}, 014012 (2011).


\bibitem{Hoodbhoy:1999dr}
  P.~Hoodbhoy, X.~Ji, and W.~Lu,
  Phys.\ Rev.\ D\ {\bf 59}, 074010 (1999).


\bibitem{Altarelli:1977zs}
  G.~Altarelli and G.~Parisi,
  Nucl.\ Phys.\ B {\bf 126}, 298 (1977).

\bibitem{Poli74} 
  H.~D.~Politzer, Phys.\ Rep.\ {\bf 14}, 129 (1974);
  D.~J.~Gross and F.~Wilczek, Phys.\ Rev.\ D {\bf 9}, 980 (1974);
  H.~Georgi and H.~D.~Politzer, {\it ibid.} {\bf 9}, 416 (1974).


\bibitem{Wakamatsu:2012ve} 
  M.~Wakamatsu,
  Phys.\ Rev.\ D {\bf 85}, 114039 (2012).


\bibitem{jizhangzhao}
  X.~Ji, J.~-H.~Zhang, and Y.~Zhao,
  in preparation.

\bibitem{Wakamatsu:2010mb}
  M.~Wakamatsu,
  Phys.\ Rev.\ D {\bf 81}, 114010 (2010).


\end{thebibliography}
\end{document}